          [2004/06/22 1.2 (KOF); 2001/04/25 1.1 (PWD)]

\documentclass[a4paper,twocolumn]{Gaia2004} % European paper size
\usepackage{times}      % for font
\usepackage{epsfig}     % for figure inclusion
\usepackage{natbib}     % for bibliography
\title{Gaia First Look}

\author{S. Jordan}
\author{U. Bastian}
\author{H. Lenhardt}
\author{H.-H. Bernstein}
\author{S. Hirte}
\affil{Astronomisches Rechen-Institut, M\"onchhofstr. 12-14,
69120 Heidelberg, Germany}
\author{M. Biermann}
\affil{Landessternwarte K\"onigstuhl, 69117 Heidelberg, Germany}

\bibpunct{(}{)}{;}{a}{}{,}  % to set bibliography punctuation to A&A style

\begin{document}


\maketitle

\begin{abstract}
A complicated and ambitious space mission like Gaia needs a careful monitoring
and evaluation of the functioning of all components of the satellite. 
This has to be performed on different time scales, by different methods,
and on different levels of precision. On the basis of housekeeping
data a Quick Look will be performed by the ground segment. A first 
analyses of the science data quality and consistency will be done 
by a Science Quick Look. However, due to the nominal scanning law a 
full self-consistent calibration of the satellite, and a determination of
astrometric and global parameters, is not possible before about half
a year has elapsed, imposing a serious danger to lose valuable observing
time if something goes wrong. Therefore, it is absolutely necessary to
perform a Detailed First Look on a daily basis on the $\mu$as accuracy level.
We describe two different methods, a block iterative 
procedure and a direct solution, to monitor all satellite parameters
that in principle can be evaluated within  a short amount of time, particularly
those that can be measured in along-scan direction. This first astrometric
analysis would greatly benefit from a modified scanning law (Zero-nu-dot
mode) for some time during the commissioning phase.

\end{abstract}

\section{Introduction}
The First-Look task aims at a rapid health monitoring of the Gaia
spacecraft and payload at the targeted level of astrometric precision.
This is difficult to achieve because
Gaia
needs a global, coherent, interleaved reduction and calibration
of about six months of measurements to reach that level of
precision. It is intended to be done by means of a Global Iterative Solution 
\citep{white-book,ll2001}, abbreviated as GIS.
Since the  GIS simultaneously solves for astrometric, attitude,
calibration, and global parameters (including relativistic
effects) one can call Gaia a self calibrating
instrument, but this can only be achieved after measurements from several
months have been gathered.
It is, on the other hand, 
of utmost importance to quickly get a handle on the
inherent quality of the elementary measurements. Learning only from
the primary science reduction that some subtle effect has degraded the
measurements would effectively mean the loss of many months of
data and mission time. It is the goal of the First-Look task to
perform an analysis of  Gaia's data on a daily basis in order to
perform the best possible monitoring of the satellite status during
the whole mission. This would allow counter-measurements if serious
problems should arise.

\section{First Look\,=\, QL, ScQL, FLP, \&\ DFL}
\begin{figure}[!t]
   \begin{center}
       \leavevmode
 \centerline{\epsfig{file=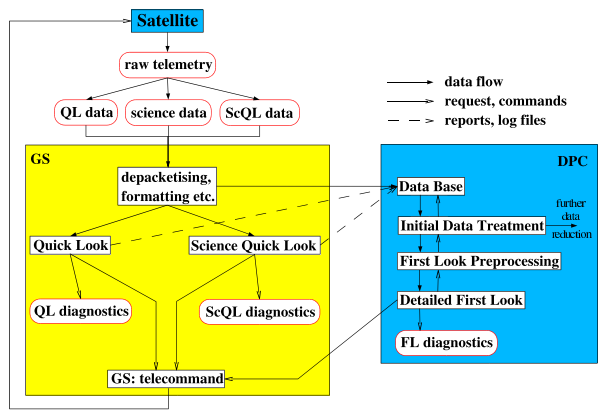, angle=270,width=1.0\linewidth}}
    \end{center}
 \caption{\label{flscheme}
Sketch to illustrate the proposed logic of the different ``Look''
tasks. The raw telemetry consists of different kinds of data (QL, ScQL and
science data) organised in different virtual channels for a simpler handling
on ground. This data is prepared for injection into the data base at the Data
and Processing Center (DPC) and for further processing. Each of the subsequent
tasks (QL, ScQL and DFL) carry out their own diagnostics to judge the bus,
instrument and science data health and (possibly) generate telecommanding and
alert the scientific Gaia consortium, respectively. Initial Data Treatment and
FLP are preparatory steps to enable the DFL task.}
\end{figure}

The monitoring of the satellite health and the quality of its data
is performed on different levels of accuracy, 
executed under different responsibilities, and with somewhat
different objectives  (see Fig.\,\ref{flscheme}).

The Quick Look (QL) task comprises all ground segment activities concerning
satellite (i.e.\ bus and instrument) health. It generates telecommanding via
the ground segment if
necessary. QL is performed in real-time or in quasi real-time and uses only
housekeeping and attitude control system data.

The Science Quick Look (ScQL) task comprises all activities concerning
scientific data health and either generates telecommanding or alerts the
scientific Gaia consortium if necessary. The Science Quick Look task will be
some less precise and simplified version of the Detailed First-Look task, intended
to immediately react on obvious deviations from what is expected for the
scientific data, such as empty windows or a very high background in the windows.
Such deficiencies are easy to detect and should quickly be handled / removed
to safe observation time.
ScQL is performed immediately after data reception on ground. 
It seems reasonable to perform it at the same location where the QL task will
be executed.

The Detailed First Look (DFL) is the in-depth scientific assessment of the
quality of the Gaia observations within about 24 hours after its reception at
the Data and Processing Center (DPC). 
At first sight this seems only possible on the basis of high-precision knowledge
of Gaia's attitude and geometric calibration, and of the positions of at least some
of the observed stars. However, as already mentioned in the introduction, a full
self-calibration Gaia can only be achieved by collecting many months of measurements.
Some compromise must thus be found to solve this apparent contradiction.

The goal of such a compromise must be a restricted astrometric calibration. sufficient
to judge the quality of the measurements but not necessarily giving absolute astrometric
and calibration parameters. We call this task First-Look Preprocessing (FLP). It is
the most complicated process in the chain. In the case of Hipparcos it was the 
strictly one-dimensional Great-Circle Reduction \citep{ESA97}.

After FLP has been performed
DFL will produce diagnostics of the
status of the satellite and instrument in a more sophisticated manner than can
be performed within QL and ScQL. The diagnostic output will be accessible in
the data base. If modifications to the satellite operations appear necessary,
this will be communicated to the ground segment.
The overall First-Look task thus consists of the sub-tasks QL, ScQL, FLP, and DFL.

First Look tasks also exist for Gaia's photometry and radial velocity
measurements. For these measurements numerous a-priori calibration objects
exist in the sky on the accuracy level of Gaia. The details of these
tasks are not described in this paper.

 \begin{figure}[!t]
  \begin{center}
    \leavevmode
 \centerline{\epsfig{file=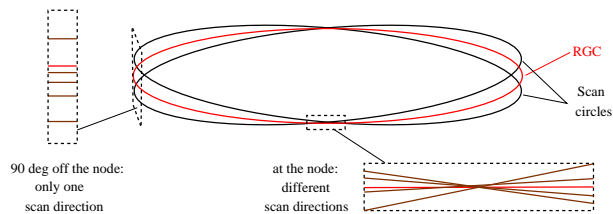, angle=270,width=1.0\linewidth}}
   \end{center}
  \caption{The RGC is defined by the scan plane in the middle
  of our one-day time interval.
  During this period the scan plane changes by only about $4^\circ$.
  Stars $90^\circ$ away from the node are moving practically parallely, 
  therefore providing no information about the $r$ coordinate perpendicular
  to the RGC. In the region around the nodes $r$ can be measured
  with limited precision.
  }
  \label{RGC}
\end{figure}

%% Sample figure environment
 \begin{figure}[!t]
  \begin{center}
    \leavevmode
 \centerline{\epsfig{file=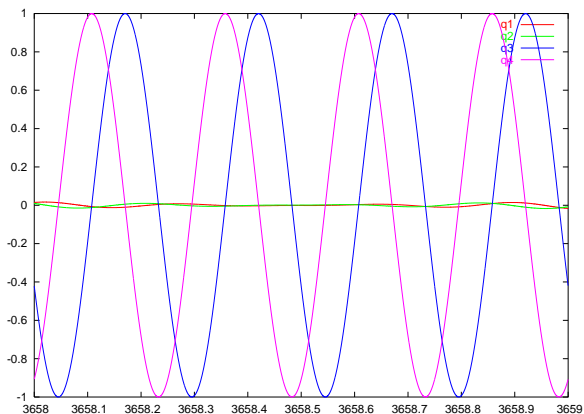, angle=270,width=1.0\linewidth}}
   \end{center}
  \caption{In the RGC system two of the quaternion components, usually
  called $q_3$ and $q_4$, have large amplitude and describe the orientation
  in $\upsilon$ direction,
  while the other two components, $q_1$ and $q_2$, 
  oscillate closely around zero and describe the slow  tilt during
  one day.}
  \label{attitude}
\end{figure}

\section{Basic Gaia data reduction}
Gaia's scanning astrometric
telescope forms optical images of celestial light sources
(mostly stars) that are
detected by TDI-operated CCDs in the focal plane.
With (at least approximately) known   geometric calibration of the telescope and focal
plane, the pixel coordinates of the centroid can be transformed into
spherical longitude and latitude coordinates. These coordinates are
called `field angles',
more specifically `observed field angles', and
denoted $\eta$~(along scan) and
$\zeta$~(across scan). For a given instant of time they
depend on 
the astrometric parameters of the source
(i.e.~barycentric celestial position at some reference epoch,
proper motion, parallax and maybe radial velocity) and  Gaia's
attitude with respect to astronomy's inertial space, defined by
the International Celestial Reference System.

In the GIS the differences between the observed field angles and
the ones computed from approximate parameters are used 
to determine corrections in a linearized least-squares
adjustment process. This is the basic principle of the self-calibrating
astrometric reduction of Gaia's measurements.

\section{First-Look Preprocessing}
The main astrometric instrument measures only along scan
(i.e.~only $\eta$). The astrometric sky mapper (ASM) measures in two
dimensions, but with a significantly lower precision 
across scan
(1-3 mas, but only, if the across-scan instrument
calibration is known, which is not the case at the beginning
of the mission). Within 24 hours
the scan plane tilts by only about $4^\circ$ degrees. A Reference
Great Circle (RGC) can be defined by the great circle perpendicular
to the direction of the rotational axis in the middle of the one-day
time interval (see Fig.\,\ref{RGC}).
The coordinate along the RGC is called $\upsilon$,
the perpendicular coordinate $r$.
$90^\circ$ away from the nodes all scans are approximately  parallel
to the RGC,
providing
only one-dimensional ($\upsilon$) 
information from the main instrument. Only in the
region around the nodes the slow tilt leads to regions where sources
are measured in slightly different directions resulting in low-precision
determinations of $r$ from multiple $\eta$ measurements. 

A Hipparcos-style great-circle reduction
won't do the First-Look job in the case of Gaia since the across-scan
position of the stars is not known with sufficient accuracy until 
the first GIS has been performed. 
The great-circle reduction could
work if the ASM were calibrated.

Whatever method we use, we cannot expect a full astrometric solution
because the problem is degenerate. Nevertheless, our goal must be 
to monitor the accuracy of the measurement on the $\mu$as level of those
parameters where it is possible. There are some aspects of the instrument and of nature that
just cannot be tackled with one day of one-dimensional measurements, which,
however, do not significantly disturb the FLP output data. Examples are a possible
along-scan `astrometric chromaticity' of the instrument, stellar proper motions
and parallaxes, and a possible
deviation of gravitational light bending from the predictions of
General Relativity.

%% Sample figure environment
 \begin{figure}[!t]
  \begin{center}
    \leavevmode
 \centerline{\epsfig{file=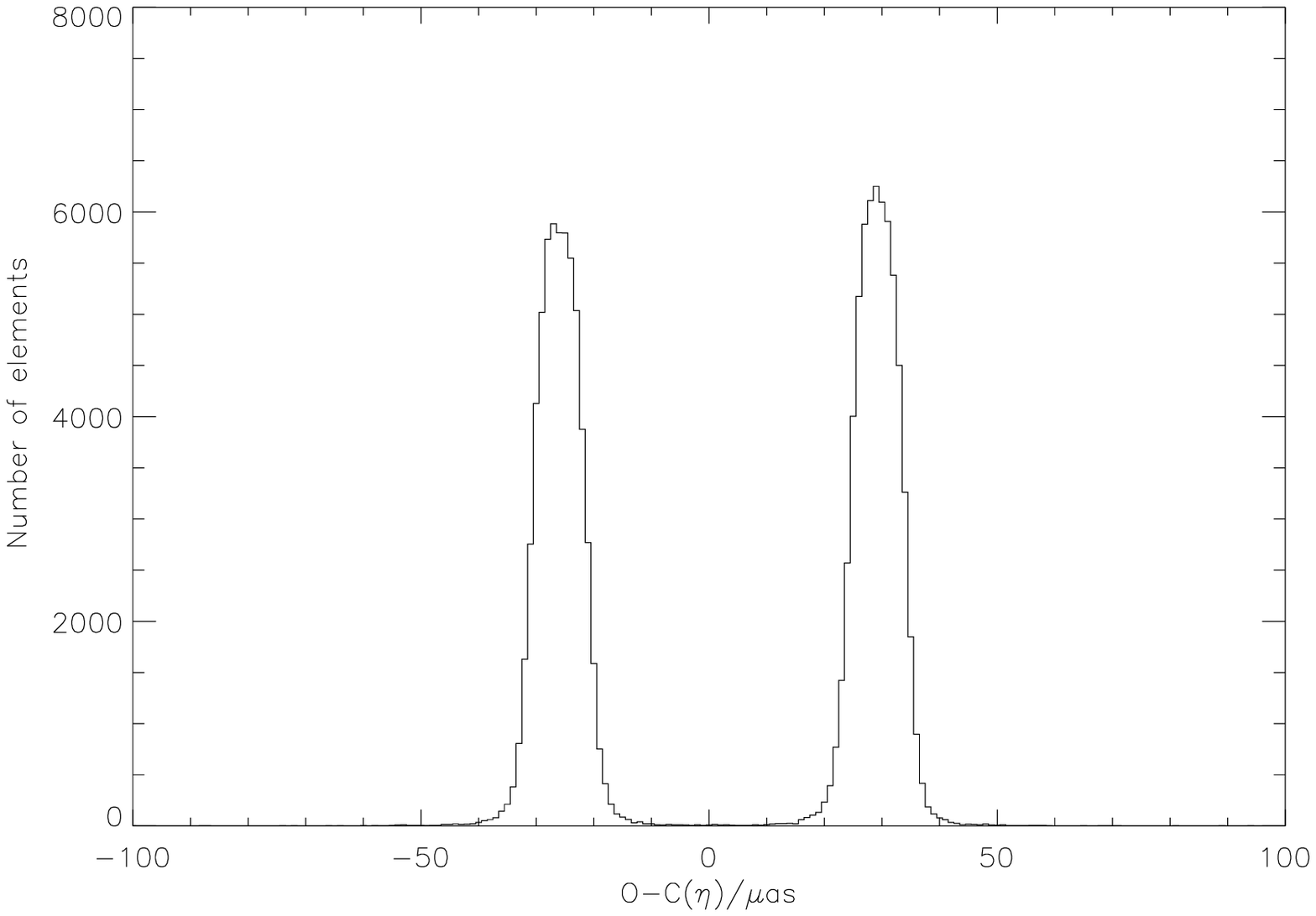, angle=0,width=1.0\linewidth}}
 \centerline{\epsfig{file=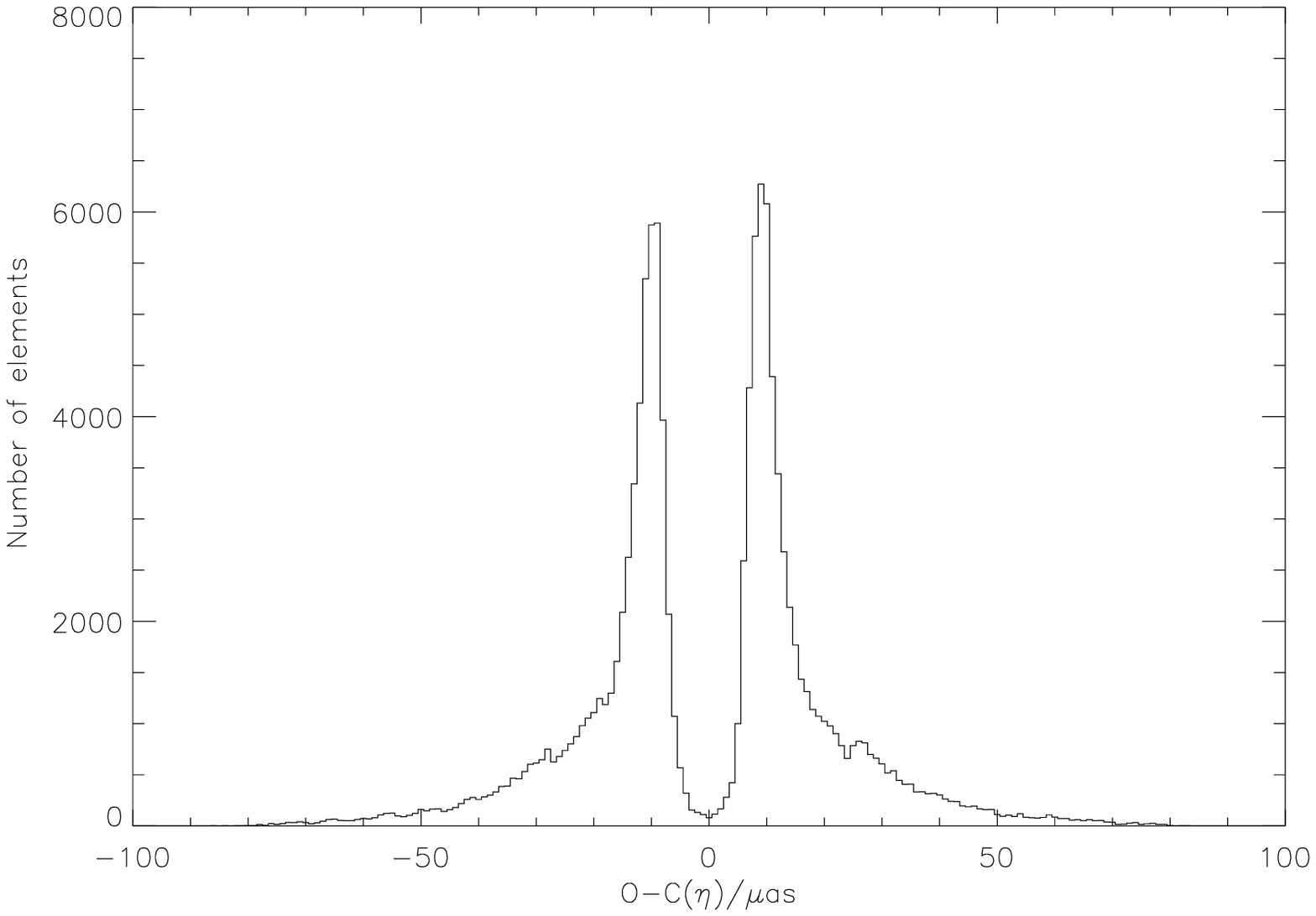, angle=0,width=1.0\linewidth}}
 \centerline{\epsfig{file=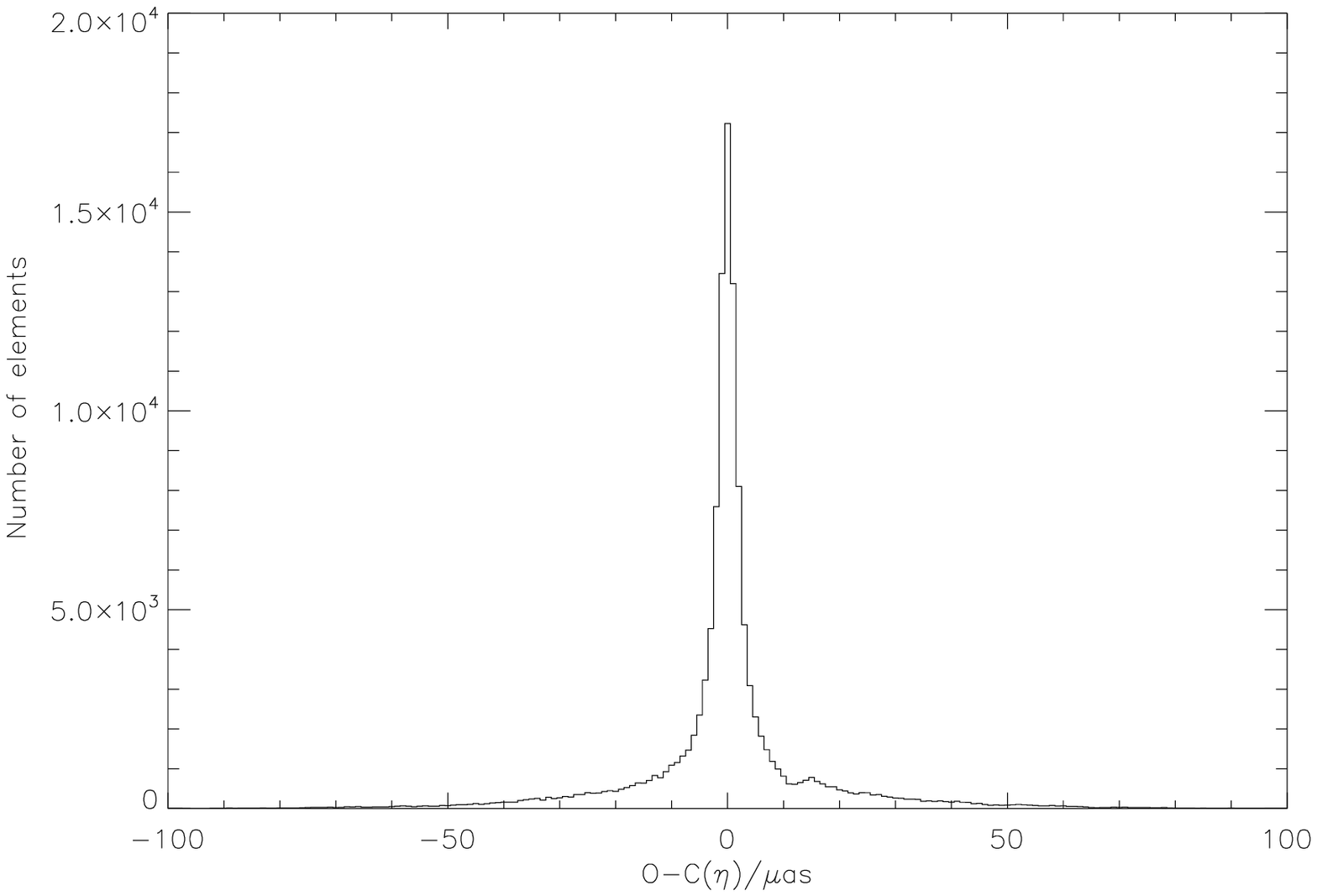, angle=0,width=1.0\linewidth}}
\end{center}
  \caption{Observed - calculated along-scan angles before the ODIS 
  iterations (top), after one iteration (center), and after 50 iterations
  (below). The double distribution originates from an assumed wrong 
  calibration angle between the two FoVs. Note that for this test data
  without any noise were used in order to test the validity of the
  simulated data and the ODIS program.
}
  \label{trueodis}
\end{figure}

%% Sample figure environment
 \begin{figure}[!t]
  \begin{center}
    \leavevmode
 \centerline{\epsfig{file=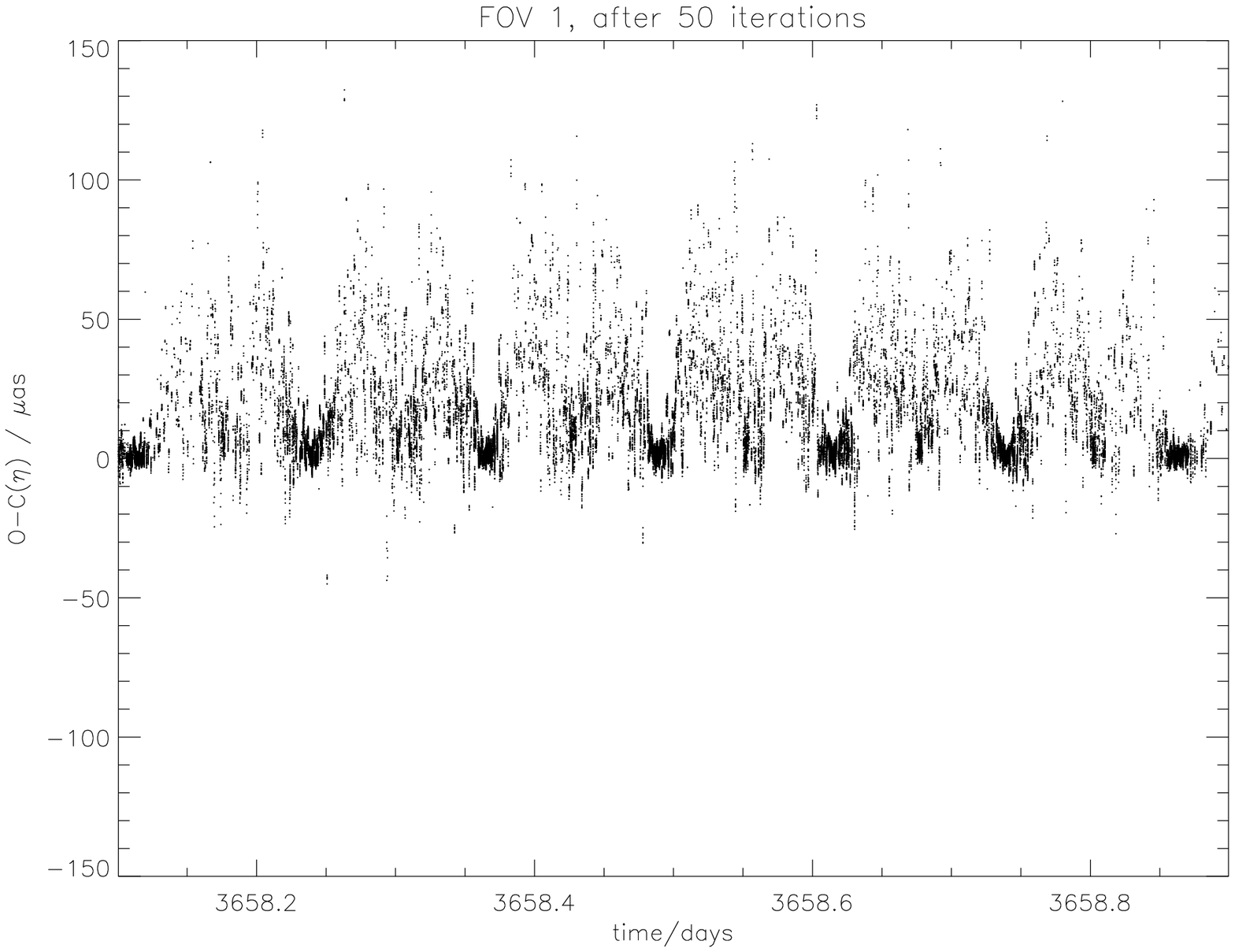, angle=0,width=1.0\linewidth}}
 \centerline{\epsfig{file=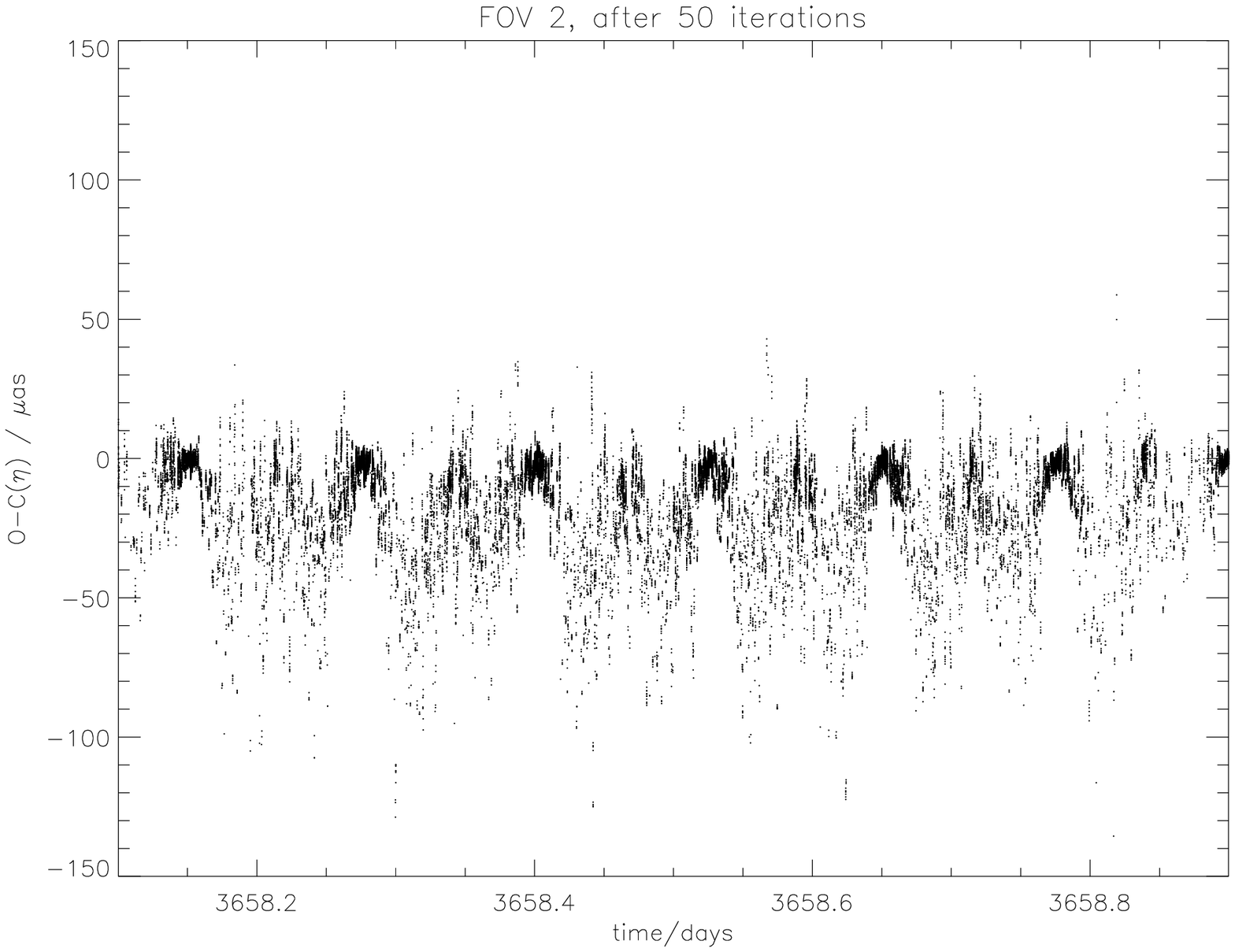, angle=0,width=1.0\linewidth}}
\end{center}
  \caption{Observed - calculated along-scan angles vs. time for
  FoV1 (top) and FoV2 (bottom) after 50 ODIS iterations.
  Note the
  unequal density  of our observations leading to strong correlations
  between the two FoVs.
}
  \label{octime}
\end{figure}

%% Sample figure environment
 \begin{figure}[!t]
  \begin{center}
    \leavevmode
 \centerline{\epsfig{file=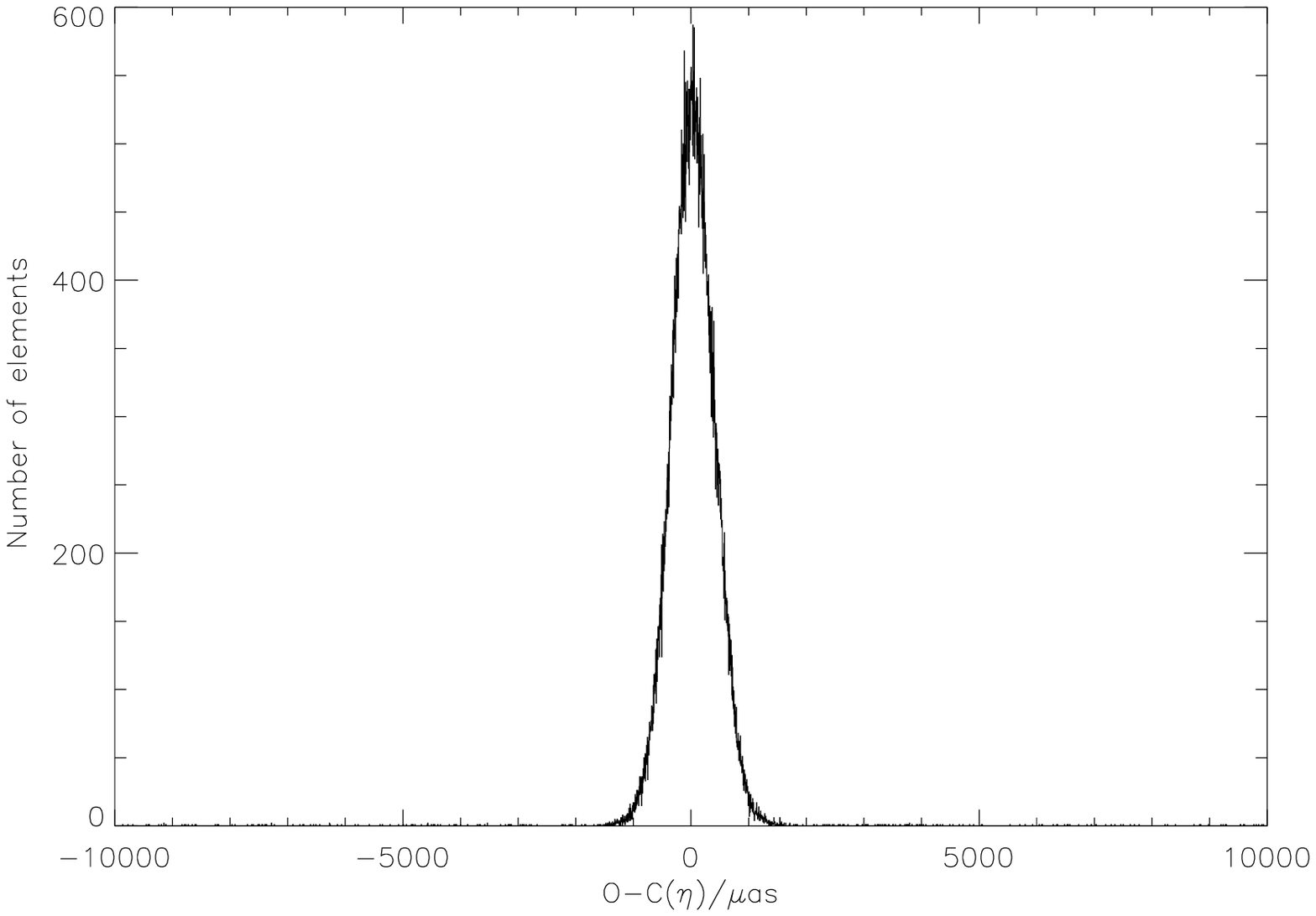, angle=0,width=1.0\linewidth}}
 \centerline{\epsfig{file=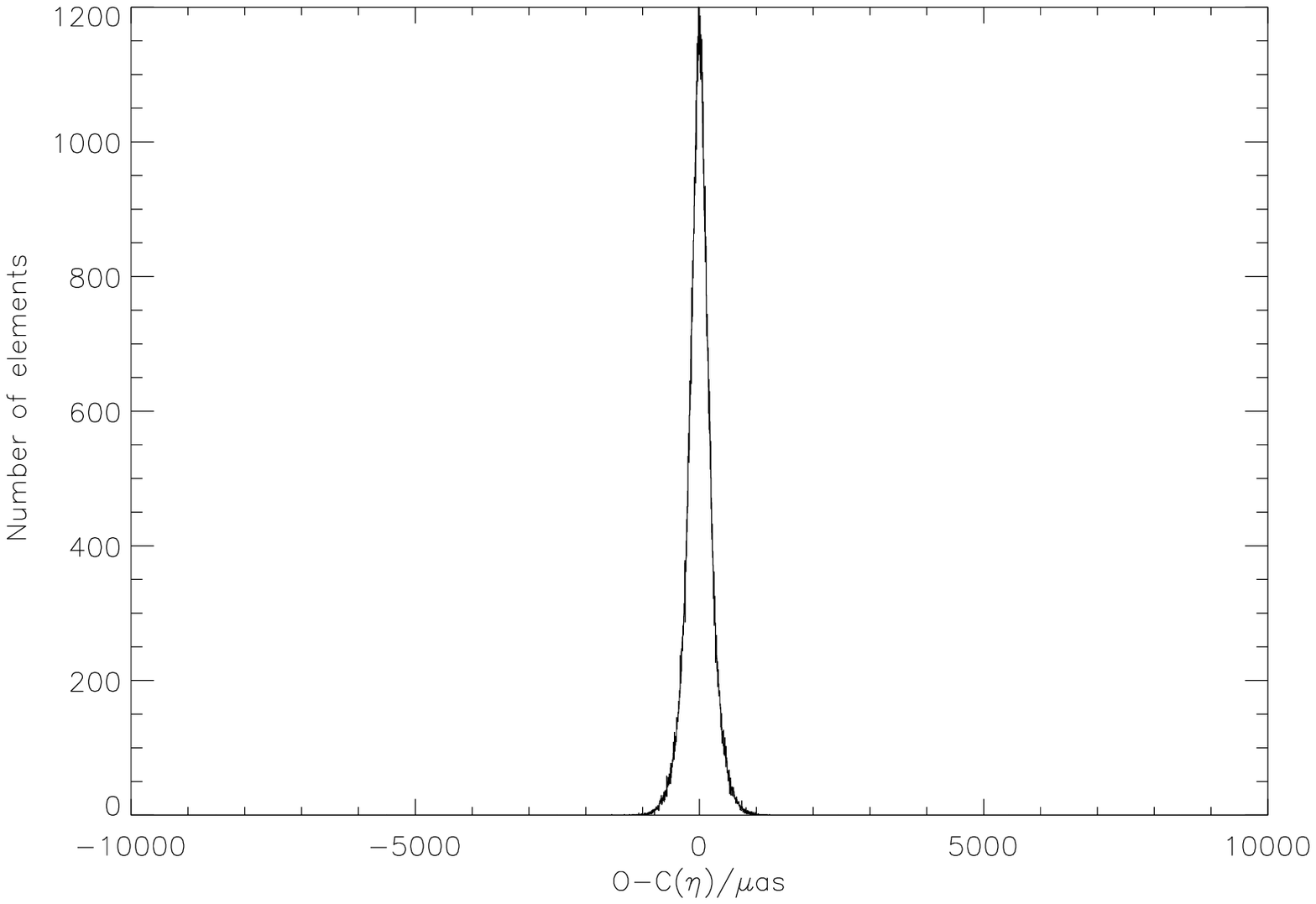, angle=0,width=1.0\linewidth}}
\end{center}
  \caption{Observed - calculated along-scan angles before the ODIS 
  iterations (top), and after 25 iterations
  (below). In this case simulated data with noise were used.
}
  \label{noiseodis}
\end{figure}

We have begun to work on two different ways to tackle the problem of
FLP: 
\begin{itemize}
\item A One-Day Iterative Solution (ODIS)\footnote{formerly called
1-day GIS, which we renamed because it is not global at all}, 
which is a reduced variant of GIS, i.e. a block-iterative method in the
spirit of GIS, but somehow identifying and omitting the degenerate
dimensions of the adjustment problem.
\item A direct setup and solution of the full
least-squares adjustment problem of FLP, using a direct elimination of the
source unknowns before inverting the remaining normal-equation matrix.
Using eigenvector decomposition of the space of unknowns allows to
identify the degenerate dimensions.
This (already quite mature) method is called `ring solution' for the
time being.
\end{itemize}

\subsection{The  One-Day Iterative Solution}
In contrast to the GIS the ODIS will not be able to solve the full astrometric
and calibrational problem due to the limitations in scan directions.
The satellite attitude will -- as in the GIS -- be described by four 
quaternions. If converted to the RGC system, two quaternion
components describe the orientation of the satellite in $\upsilon$ direction, 
while the other two account for the slow tilt during the 24 hour
time interval (see Fig.\ref{attitude}, see \cite{sj-002}).
In the current version the fitting parameters for
the attitude are 1440 quaternion (for 24 hours)
B-spline coefficients for the quaternions 
at equidistant nodes
every 60 seconds; it may be necessary to use a higher density of nodes.
1440 quaternions translate into 5760 unknowns for the four components
of which one quarter is redundant due to the constraint that
the quaternions have length unity. This constraint is accounted for
by pseudo observing equations.

The CCD calibration can be described by 3 along-scan unknowns per CCD
(e.g. `shift', `rotation' and `distortion') leading
to  510~unknowns. We estimate the number of observed sources
during one day to be about 3 million on average. Since we cannot try
to measure more than $\upsilon$ and a lower number
of $r$ coordinates, the number of source unknowns is less than 
6 million.
This number will be further reduced if we consider only those sources 
observed at least 2 times.
The number of elementary measurements of $\eta$ and $\zeta$ is
of the order of  50 million.

For our software development and first tests we use 
24 hours of data 
provided to us by the   Gaia Simulation Working Group, calculated
in the framework of the 
Gaia-1 instrument model --  which is no important
restriction for our tests and will be adapted to the current model in the
future.

For a first consistency test we used the `true' data, i.e. without any noise
in attitude, source positions, and elementary measurements. For the
latter we assumed that we can also precisely measure the across-scan
angle $\zeta$. Furthermore, we limited  ourselves to 4\%\ of the
measurements from only one of the 16 strips of CCDs and took into account only
those sources observed at least three times.
This amounted to 140\,000 measurements of 30\,000 sources.
While we began with `correct' starting values for all other
unknowns, we assumed a somewhat wrong
along-scan calibration of the CCDs. 
The top panel of Fig.\,\ref{trueodis} shows the distribution of the
observed minus calculated along-scan angles $\eta$ before the first
ODIS iteration. It is clearly visible that the observations from both
fields of view (FoVs) do not fit together and result in a bimodal
distribution. After the first iteration (center panel),
both distributions came closer
together, and after 50 iterations we have a single distribution
with a sharp peak of only 3 $\mu$as width (bottom).
However, besides the sharp peak we also find a broad background of
strongly deviating $\eta$ values. 
In Fig.\,\ref{octime} we plot the 
observed minus calculated $\eta$ vs.\ time for both FoVs and see that
the strongest deviations from zero occur where the density of
elementary measurements is small.
This reflects the varying density with galactic latitude 
amplified by our negligence of those
stars which have been observed less than three times.
Therefore, while one
FoV may be in a region of low density of considered sources the other
may be in a high density region, giving very unequal weight. This makes
it impossible for the attitude parameters to converge and leads to
the 
broad background in Fig.\,\ref{trueodis}. We expect that this effect
is strongly reduced if a higher number of observations are taken into
account and if we require only one repeted observation of every
source.

However, our simulation clearly shows that the vast majority of 
the $\eta$s is concentrated around zero in the observed minus computed
diagram after the iterations so that we conclude that our software works properly.

In a next step we repeated the same procedure with Gaussian noise
of $0.1$\,mas in along-scan and $2$\,mas in across-scan direction.
Noise was also present in the initial approximation of the
attitude. 
The latter value would be a realistic error estimation if we had
a relatively precise calibration of the ASMs. Note, that this sort
of calibration cannot be performed within the ODIS but would
-- before the first GIS -- need a separate calibration step 
(see Sect.\,\ref{zeronu}). Differently from the case of our simulation
with the `true' data, we started with the `correct' CCD calibration.
The top panel of Fig.\,\ref{noiseodis} shows the deviations of the
observed from the calculated $\eta$s before the first iteration,
the lower panel the distribution after 26 iterations. The width of
distribution is consistent with the assumed accuracy of the data
and an average number of observations per source of 4. 
Due to the large total number of observations, even small deviations
of this width from the expected one will allow a diagnosis of possible
problems in calibration and attitude, particularly if further 
investigated in time (as Fig.\,\ref{octime}). 

The ODIS by itself will not directly show the degeneracy of the 
problem. Another disadvantage is that there is no guarantee that
the block-iterative scheme will converge to the correct solution.

By investigating a simple partially degenerate problem, 
\cite{sj-001} could show that the convergence behavior strongly depends
on the  block sequence, on block size, and on the block composition.
It turned out that it would be an advantage  to split up all three blocks
(for attitude, sources, calibration; global parameters cannot be
taken into account in ODIS so that the global
block is omitted) into
six blocks solving for along and across-scan, or more
precisely, in direction of the RGC and perpendicular direction, separately.
This would extremely accelerate the convergence and minimize
the contamination of the solution of the well defined along-scan
unknowns
by the ill-conditioned across-scan unknowns.
The attitude block, i.e. will be split up into a block for the
quaternion components $q_1$ and $q_2$ for the across-scan directions,
while the block for $q_3$ and $q_4$ would solve for the along-scan
direction.

The next steps will be to test what kind of information can be gathered
if we assume that no across-scan information from the ASMs is available
(which would be the case if no across-scan calibration of the CCD is
possible). We will also make use of the full data set of simulated
data. 

By assuming that the CPU time scales linearly with the number of
elementary measurements we can already estimate the amount of
data and the computational time for such a longer run with 300 times
more observations: Each iteration will need about 10 hours on a 
2 GHz Pentium III. When optimized, four ODIS iterations should be sufficient.
Since the solutions for the astrometric parameters for each star
are independent from each other the major part of the ODIS can 
easily be  parallelized. Thus ODIS seems easily feasible computationally.

\subsection{The Ring Solution}
The limited number of unknowns during one day makes it possible to try a
direct solution of the problem rather than an iterative solution. 
Such an approach makes it possible to further analyze the degeneracy of
the problem in more detail and allow diagnostics which are not available
in the case of iterative methods.

After initializing counters, matrices, {\em a~priori} approximations
of attitude and calibration, the process starts a loop over the
relevant observations, in a source-by-source sequence.
All available observation equations of a given source
are set up and accumulated into
normal equations (i.e.\ normal-equation matrix and right-hand side vector).

Next we analyze every  submatrix for the source unknowns separately
in order to find out whether it is degenerate due to insufficient
across-scan information. If this is the case, only the along-scan unknowns
are retained. We will also investigate analytically whether all sources having
only one field-of-view transit must be eliminated.

The remaining non-degenerate source part is then directly eliminated
from the entire normal-equation contribution of this source, using
a Gauss elimination scheme. This leaves a reduced system of
normal equations for the calibration and attitude unknowns only.

These reduced equations are accumulated in a source-by-source fashion,
until the complete normal equations for the reduced problem
(calibration and attitude unknowns only) have been set up.

The well-known, obvious degeneracies of any global astrometric
adjustment (rigid rotation of the adjusted system with respect to
inertial, astronomical space) are eliminated by adding simple,
well-demonstrated constraints.

The eigenvalue spectrum allows to judge the condition of the system
and restrict it to the non-degenerate dimensions. The restricted system
is solved, using a Moore-Penrose pseudo-inverse of the normal-equation matrix.

Afterwards, the reduced solution is back-substituted into the source
parts, yielding the source unknowns and --- finally --- the residues of the
individual measurements.

 \begin{figure}[!t]
  \begin{center}
    \leavevmode
 \centerline{\epsfig{file=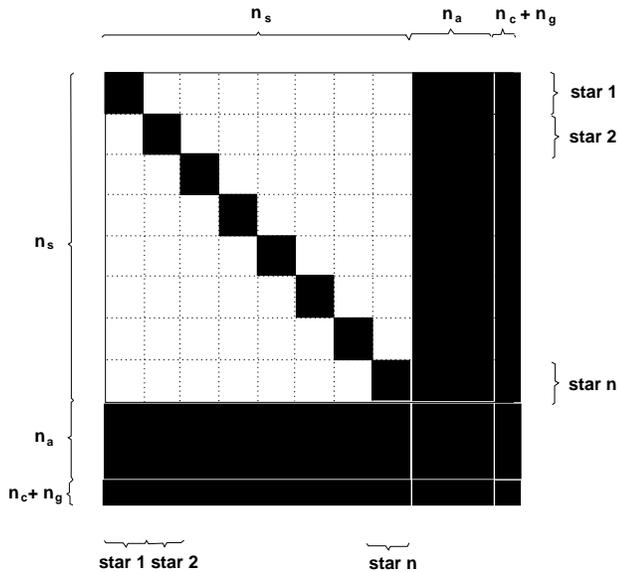, angle=270,width=1.5\linewidth}}
   \end{center}
  \caption{Structure of the normal equation matrix.
  $n_a$, $n_s$, $n_c$, and $n_g$ denote the number of
  attitude, source, calibrational, and global unknowns. 
  The submatrix for the source unknowns can be divided
  into small independent blocks allowing a star-by-star
  solution after the large blocks for the attitude, and calibration
  parameters has been evaluated. $n_g$ is zero for FLP.}
  \label{matrix}
\end{figure}

The whole procedure can be iterated a few times, for various purposes:
\begin{itemize}
\item to get rid of the (small) non-linearities in the problem
\item to omit discordant individual measurements (`outliers')
\item to omit disturbed sources (unresolved doubles, variables,
     extended objects, ...)
\end{itemize}

The basic structure of the normal equations is seen in
Fig.\,\ref{matrix}. The largest block that has to be solved as
one block is a matrix containing a few thousand attitude B-spline
coefficients and a few hundred calibration parameters.

In the near future we will begin numerical tests with the Ring Solution
with the same input data that we used for the ODIS. We can then compare
the strengths and shortcomings of both methods in terms of computing time,
memory demand, and possibilities to build up First-Look 
monitoring tools for the DFL. We expect that ODIS will be much faster,
but particularly a detailed study of the rank deficiencies should
be possible only with the Ring Solution. Nevertheless, the two methods will 
complement each other so that both paths should be followed.

Our current estimate on  a 2 GHz Pentium III system 
is that about 800 hours of CPU time are needed if the procedure is
iterated two times for the whole 24 hour data set. This number is
actually not alarmingly high because we believe that several optimizations
in the numerical scheme and the applied methods are possible. Moreover,
at the time of Gaia much faster computers will be available. Many parts
of the Ring Solution can also be parallelized.

More details of the Ring Solution can be found in 
\cite{bst-001}.

In principle one could consider a direct method similar to the
Ring Solution as an alternative
to the GIS. It has, however, to be analyzed whether the much larger 
number of attitude unknowns will allow such a solution with reasonable
computational effort. Since we do not expect degeneracy problems
as serious as in the Ring Solution for a single day, one would not 
so much depend on the knowledge of the inverse of the normal equations.
Therefore, more efficient mathematical methods to solve such equations
without inversion, like Krylov subspace methods may allow the
application of direct methods for the astrometric solution of the
Gaia data.

\section{The Detailed First Look}
     
The goal of both FLP methods is to find out whether the measurements lie within
the predicted error range so that the mission's objectives can be
reached.
The basic procedure to achieve this goal is an in-depth investigation
of the astrometric residuals after FLP, and of the astrometric unknowns derived
from FLP and their variances and covariances, respectively.
A careful analysis of the time dependence of the
residuals by means of power spectra, wavelet analysis, or filtering is
important. If periodic or other correlated oscillations occur, they
can be used to test or update the mechanical model of the satellite.
It is also planned to search for a dependence of the residuals on
object classes with the help of auto-correlation functions, e.g.\ with
respect to brightness and colour. 
In order to understand the  details (e.g.\ mean, variance, and 
kurtosis of $\chi^2$ distribution functions, auto-correlations)
of the observed data, simulations under various physical hypotheses will
be performed. Sophisticated hypothesis testing methods will then allow
a detailed diagnostic of the behaviour of the satellite.

\section{Calibration and Zero-nu-dot Mode}
\label{zeronu}
A basic problem of any kind of 24-hour astrometric solution is the
lack of across-scan information. Even a limited knowledge in $r$ 
direction would extremely reduce the degeneracy of the problem.

Prerequisite for such a measurement would be an accurate calibration of
the ASMs. Since no ground-based calibration would be exact enough this
has to be done in space. Ideally one would use a
large number of observations of the same
stars with well known positions in as many as possible scan directions.

\begin{figure}[!t]
  \begin{center}
    \leavevmode
 \centerline{\epsfig{file=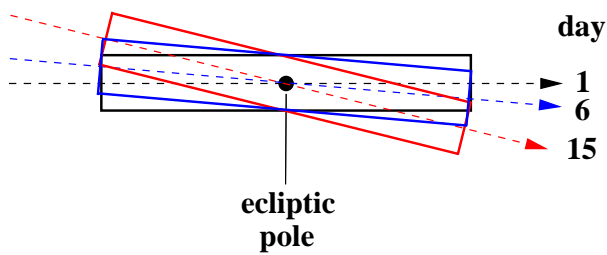, angle=270,width=1.0\linewidth}}
   \end{center}
  \caption{In the zero-nu-dot mode the satellite axis remains on the ecliptic
  $50^\circ$ away from the sun. Therefore, the ecliptic poles are the
  nodes of this scanning law. The scan direction changes by about
  $1^\circ$ per day so that the stars around the ecliptic pole
  are measured in increasingly different directions.}
  \label{zeronudot}
\end{figure}

The normal Gaia scanning law is quite inappropriate for this purpose 
since the nodes of the scanning law are moving too fast to allow 
multi-direction measurement of the same ensemble of stars in a short
time interval. For this reason, a somewhat modified scanning law would
help for the initial calibration: We propose that  at the beginning of the mission
or during the commissioning phase on the
transfer to L2 the scan axis should remain
on the ecliptic preceding or following the sun at the nominal $50^\circ$ angle.
The  precessing revolving angle $\nu$ of the
nominal scanning law will remain constant in this phase. This is
why we call this the zero-nu-dot ($\dot{\nu}$) mode. $\nu$ itself
would be either $0^\circ$ or $180^\circ$. The nodes of this scanning
law are the two ecliptic poles. Since the sun moves by about $1^\circ$ per
day, the stars directly at the poles will be observed under increasingly
different scan directions. Within 15 days 
we have already measurements differing up to $15^\circ$ (see Fig.\,\ref{zeronudot}).

This mode could be prepared and augmented complemented by high-precision
ground-based observations of a few thousand stars around each of the
ecliptic poles, aiming at {\em relative} astrometry with a precision of
a few mas, performed shortly (a year or so) before the launch
of Gaia. 
This would give an instant in-orbit astrometric across-scan calibration
of Gaia at the mas level, even without FLP, i.e. on the ScQL level.
If we limit the pre-launch astrometry to two circles of $1^\circ$ 
radius around each pole we expect about 8000 stars
with  $V<16$ at this galactic latitude \citep{allen}. This is the
minimum, an optimum would be twice the radius, i.e. four times as many stars.
Besides an improved calibration we would also obtain a first
small ensemble of stars at the ecliptic poles with positional
improvement from Gaia.
 
\section*{Conclusions}
We have demonstrated the need of a First Look task for the Gaia mission. In particular the
DFL on an astrometric level performed on a daily basis will significantly reduce the
risk of losing valuable observing time due to problems which would otherwise become apparent
after about half a year of data collection. The necessary FLP is by no means trivial due to
much larger accuracy of the measurement along the RGC compared to the
perpendicular direction. This drawback can partially be reduced when an in-flight 
calibration of the ASMs during the commissioning phase can be performed, e.g.\ with the help
of a zero-nu-dot scanning mode. 

The major and most complicated part of the First-Look task  consists of the FLP which can be performed either
by a block-iterative procedure (ODIS) or a direct (Ring) solution. Our first computational tests
show that the ODIS will be computationally easy. With the help of the Ring solution we expect to
develop more sophisticated diagnostic tools and have estimated that this method  -- although more CPU time
demanding than the ODIS -- will also be feasible with affordable computational effort.

\section*{Acknowledgments}
We are grateful to
Lennart Lindegren for providing us FORTRAN routines
for the GIS (LL37.f). We also thank Eduard Masana and his collaborators
for producing the tailor-made simulated data for a 24-hour period.

\end{document}